\documentclass[amsmath,amssymb,aps,superscriptaddress,twocolumn,mathtools]{revtex4-2}
\usepackage{graphicx}
\usepackage{float}
\usepackage{dcolumn}
\usepackage{bm}
\usepackage[utf8]{inputenc}
\usepackage{natbib}
\usepackage{xcolor}
\usepackage{svg}

\begin{document}

\preprint{APS/123-QED}

\title{Microwave Spin-Pumping from an Antiferromagnet FeBO$_3$}
\author{D.A.Gabrielyan}
\thanks{Authors to whom correspondence should be addressed:\\ \url{davidgabrielyan1997@gmail.com}}
\affiliation{
	Kotel'nikov Institute of Radioengineering and Electronics, Russian Academy of Sciences, Moscow 125009, Russia
}%
\affiliation{
	Moscow Power Engineering Institute, Moscow 111250, Russia
}%
\author{D.A. Volkov}
\affiliation{
	Kotel'nikov Institute of Radioengineering and Electronics, Russian Academy of Sciences, Moscow 125009, Russia
}%
\affiliation{
	Moscow Power Engineering Institute, Moscow 111250, Russia
}%
\author{E.E. Kozlova}
\thanks{\url{elizabethkozlova1@gmail.com}}
\affiliation{
	Kotel'nikov Institute of Radioengineering and Electronics, Russian Academy of Sciences, Moscow 125009, Russia
}%
\affiliation{
	Moscow Institute of Physics and Technology, Dolgoprudny 141701, Russia
}%
\author{A.R. Safin}
\affiliation{
	Kotel'nikov Institute of Radioengineering and Electronics, Russian Academy of Sciences, Moscow 125009, Russia
}%
\affiliation{
	Moscow Power Engineering Institute, Moscow 111250, Russia
}%
\author{D.V. Kalyabin}
\affiliation{
	Kotel'nikov Institute of Radioengineering and Electronics, Russian Academy of Sciences, Moscow 125009, Russia
}%
\affiliation{
	HSE University, Moscow, 101000, Russia
}%
\author{A.A. Klimov}
\affiliation{
	Kotel'nikov Institute of Radioengineering and Electronics, Russian Academy of Sciences, Moscow 125009, Russia
}%
\affiliation{
	MIREA - Russian Technological University, Moscow 119454, Russia
}%
\author{\\ V.L. Preobrazhensky}
\affiliation{
	Prokhorov General Physics Institute RAS, Moscow 119991, Russia
}%
\author{M.B. Strugatsky}
\affiliation{
	V.I. Vernadsky Crimean Federal University, Simferopol 295007, Russia
}%
\author{S.V. Yagupov}
\affiliation{
	V.I. Vernadsky Crimean Federal University, Simferopol 295007, Russia
}%
\author{I.E. Moskal}
\affiliation{
	Kotel'nikov Institute of Radioengineering and Electronics, Russian Academy of Sciences, Moscow 125009, Russia
}%
\author{G.A. Ovsyannikov}
\affiliation{
	Kotel'nikov Institute of Radioengineering and Electronics, Russian Academy of Sciences, Moscow 125009, Russia
}%
\author{S.A. Nikitov}
\affiliation{
	Kotel'nikov Institute of Radioengineering and Electronics, Russian Academy of Sciences, Moscow 125009, Russia
}%
\affiliation{
	Moscow Institute of Physics and Technology, Dolgoprudny 141701, Russia
}%
\affiliation{
	Saratov State University, Saratov 410012, Russia
}%

\begin{abstract}

 Recently, canted antiferromagnets offer great potential for fundamental research and applications due to their unique properties. The presence of the Dzyaloshinskii-Moriya interaction leads to the existence of a weak ferromagnetic moment at room temperature. We study both theoretically and experimentally microwave spin pumping by the quasi-ferromagnetic mode from a canted easy plane antiferromagnet with weak ferromagnetism FeBO$_3$. The conversion of a microwave signal into the constant voltage is realized using the inverse spin Hall effect in an iron borate/heavy metal heterostructure. We use an additional bias magnetic field to selectively tune the resonance frequency of such a microwave detector over a wide range up to 43.5 GHz with potential sensitivity near 2.5 $\mu$V/W. We confirm the pure spin current nature by changing polarity of the detected via inverse spin Hall effect voltage by switching the direction of the bias magnetic field. We believe that our results will be useful for the development of highly tunable, portable and sensitive microwave antiferromagnet-based functional devices. 
\end{abstract}

\maketitle


Spin pumping in magnetic materials is one of the most powerful methods for detecting pure spin currents. Spin currents in ferromagnetic metals were the primary focus of spin pumping research\,\cite{Saitoh2006, Azevedo2005}, but these studies were gradually extended to magnetic semiconductors and isolators\,\cite{Heinrich2011, Takahashi2016}. Researchers in modern spintronics are focused on multisublattice magnetic materials such as antiferromagnets (AFM), because due to exchange enhancement, they exhibit a number of unique properties\,\cite{Xiong2022, Baltz2018, Jungwirth2016}: subterahertz and teraherz (THz) resonance frequencies, high speed of spin waves and domain walls, amplification of spin current due to excitation of evanescent modes, etc. These materials are the most promising candidates for applications in the field of ultrafast magnonics and spintronics\,\cite{Rezende2019, Barman2021} such as the new generation of 5G and 6G telecommunications. This is also supported by low radiation losses to AFM, meaning that they create virtually no electromagnetic microwave scattering fields due to zero total magnetic moment. Various AFM-based functional devices have been described in detail previously: oscillators\,\cite{Cheng2016, RomanKhymyn2017, Sulymenko2017, Popov2020, Consolo2021}, detectors\,\cite{Khymyn2017,Gomonay2018,Safin2020,Safin2022,Kozlova2022}, emitters and amplifiers\,\cite{Stremoukhov2019,Khymyn2016}, neuromorphic processors\,\cite{Bradley2023,Sulymenko2018,Mitrofanova2022}, memory elements\,\cite{Kosub2017,Fina2020}, spectrum analyzers\,\cite{Artemchuk2020,Mitrofanova2023} etc.

Spin pumping experiments have been conducted with various AFM materials, namely Cr$_2$O$_3$\,\cite{Li2020}, MnF$_2$\,\cite{Vaidya2020,Ross2015} and NiO\,\cite{Stremoukhov2022}. However, the above-mentioned materials have resonance frequencies from hundreds of gigahertz (GHz) to units of terahertz, which significantly complicates the experimental measurement of resonance characteristics by detecting spin currents. But there are AFM materials with weak ferromagnetism caused mostly by the Dzyaloshinskii-Moriya interaction (DMI), having resonance frequencies as in several hundred GHz (''optical'' antiferromagnetic modes), and in units and tens of GHz (''acoustic'' quasi-ferromagnetic modes). It opens up new opportunities for experimental research of spin currents in AFM. The generation of spin currents in various canted multisublattices magnets is a crucial and weakly explored field. Hematite, $\alpha$-Fe$_2$O$_3$, is an AFM with weak ferromagnetism, it is an excellent candidate for these investigations. Recently\,\cite{Boventer2021,Wang2021,Lebrun2020}, it has been published the spin-pumping from the hematite on the lower frequency mode across the Morin phase transition.

\begin{figure*}[t]  
\includegraphics[width=1\linewidth]{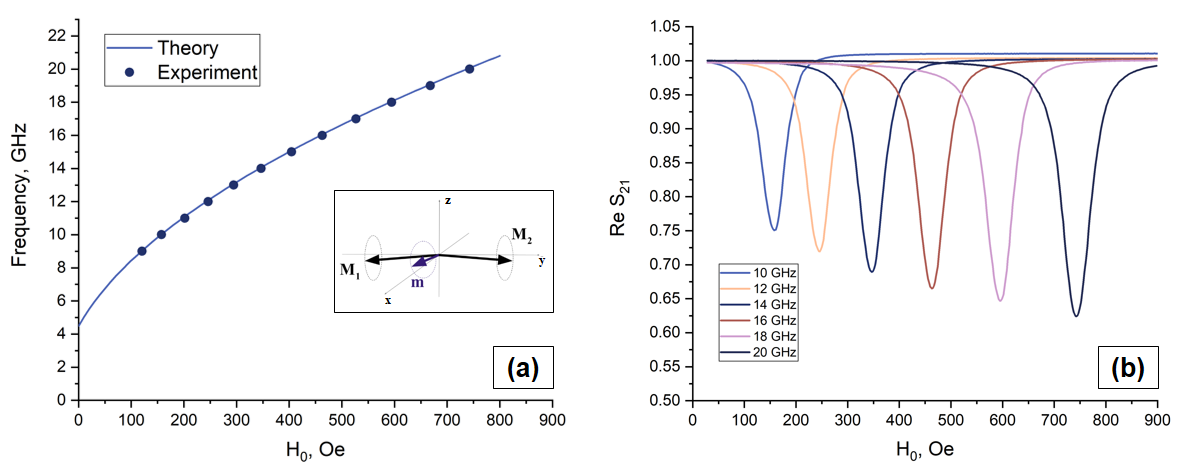}
\caption{
(a) Theoretical and experimental dependences measured by the \textit{VNA-FMR} method magnetic field of the resonant frequency of the FeBO$_3$. The insert shows the precession of the magnetic sublattices $\mathbf{M}_{1,2}$ in the antiferromagnet and of the magnetization vector $\mathbf{m}$. (b) FMR-spectra of the FeBO$_3$ at various frequencies measured by the \textit{VNA-FMR} method. Under the action of the external magnetic field, the precession of the total magnetization vector of the sublattices in the antiferromagnet is occurred.  
}
\label{fig:01}
\end{figure*}

In this work, we study both theoretically and experimentally microwave spin pumping from the iron borate/platinum heterostructure. Iron borate (FeBO$_3$) is an antiferromagnet with weak ferromagnetism and easy plane magnetic anisotropy. It has a trigonal (rhombohedral) crystal structure, a high Neel temperature (348 K) and narrow lines of magnetic resonances \cite{Turov2004, Dmitrienko2014, Kuzmenko2014, Afanasiev2014, LVVelikov1974, Jantz1978, Diehl1975, Ovchinnikov2020}. Iron borate crystals have two magnetic sublattices. The magnetic moments of their ions are located at a small angle of 55$^\prime$ \cite{Petrov1972} caused by the Dzyaloshinskii-Moriya interaction \cite{Ovchinnikov2020, Zvezdin2020}. Along with the existence of a non-zero antiferromagnetic moment, there is also a weak ferromagnetic moment $4 \pi M$ = 115-120 G at room temperature \cite{Kadomtseva1972}. In this work we use a highly advanced single-crystalline iron borate samples synthesized by a flux growth technique\,\cite{Yagupov2018}. FeBO$_3$ has two modes of its eigen-oscillations - quasi-ferromagnetic (tens of GHz) \cite{LVVelikov1974, Jantz1978} and quasi-antiferromagnetic (hundreds of GHz) \cite{Mashkovich2019}. All the above makes iron borate a promising material for fundamental research, as well as for creating various magnetoelectronics devices on its basis.  

We measure the spin current using the FeBO$_3$/Pt heterostructure for the first time, and also compare the results with a theoretical model, taking into account the Dzyaloshinskii-Moriya interaction and the magnetoelastic coupling. The analysis demonstrates the important role of these components in describing antiferromagnetic processes with weak ferromagnetism. 

We start with the theoretical and experimental investigation of the low-frequency mode and S$_{21}$-parameter, which gives information about the absorbed power of the material. The insert in Fig.\,\ref{fig:01}(a) demonstrates that a homogeneous precession of the magnetic sublattices $\mathbf{M}_{1,2}$ in antiferromagnet is initiated by the action of an external variable microwave radiation with linear polarization. The dynamics of the magnetization vector $\mathbf{m}=\left(\mathbf{M}_{1}+\mathbf{M}_{2}\right)/2\textrm{M}_{\tiny{\textrm{s}}}$ and the N\'eel vector $\mathbf{l}=\left(\mathbf{M}_{1}-\mathbf{M}_{2}\right)/2\textrm{M}_{\tiny{\textrm{s}}}$ ($\textrm{M}_{\tiny{\textrm{s}}}$ is the saturation magnetization of the AFM) are governed by the equations, respectively\,\cite{ozhogin1977effective,zvezdin1979dynamics,bar1980nonlinear,Andreev1980}:

\begin{equation}\label{eq:01}
\begin{array}{c}
    \mathbf{m}=\frac{1}{\omega_{\tiny{\textrm{ex}}}}\left(\left[\frac{d\mathbf{l}}{dt}\times\mathbf{l}\right]+\left[\mathbf{l}\times\gamma\left(\mathbf{H}_0+\mathbf{h}_{\tiny{\textrm{RF}}}\right)\right]\times\mathbf{l}+\left[\mathbf{l}\times\gamma\mathbf{H}_{\tiny{\textrm{DMI}}}\right]\right),
    \end{array}
\end{equation}
\begin{equation}\label{eq:02}
\begin{array}{c}
    \mathbf{l}\times\left[\frac{d^2\mathbf{l}}{dt^2}+\gamma_{\tiny{\textrm{eff}}}\frac{d\mathbf{l}}{dt}+\frac{\partial{W}}{\partial{\mathbf{l}}}-2\gamma\left[\frac{d\mathbf{l}}{dt}\times \mathbf{H}\right]\right]\\
    =\gamma\left[\frac{d\mathbf{H}}{dt}\times \mathbf{l}\right],
    \end{array}
\end{equation}
where $\gamma_{\tiny{\textrm{eff}}}=\alpha_{\tiny{\textrm{eff}}}\omega_{\tiny{\textrm{ex}}}$ is the spectral linewidth of the AFM resonance, $\alpha_{\tiny{\textrm{eff}}}$ is the effective damping including Gilbert constant and spin-pumping terms (see for more details\,\cite{Baltz2018,Cheng2014}), $\omega_{\tiny{\textrm{ex}}}$ is the exchange frequency, $\gamma=2\pi \cdot 28$ GHz/T is the gyromagnetic ratio, $\mathbf{H}=\mathbf{H_0}+\mathbf{H_{\tiny{\textrm{DMI}}}}+\mathbf{h_{\tiny{\textrm{RF}}}}$, here $\mathbf{H_0}=\textrm{H}_0\mathbf{x}$ is the external constant magnetic field, $\mathbf{H_{\tiny{\textrm{DMI}}}}=\textrm{H}_{\tiny{\textrm{DMI}}}\mathbf{z}$ is the DMI field and the magnetic field component of a linearly polarized microwave field $\mathbf{h_{\tiny{\textrm{RF}}}}=\textrm{h}_{\tiny{\textrm{RF}}}\cos{\omega \textrm{t}}\cdot\mathbf{y}$, $W(\mathbf{l,H_0})$ is the potential function \cite{Ozhogin1988, Safin2022}:
\begin{equation}\label{eq:03}
\begin{array}{c}
    W(\mathbf{l,H_0})=\frac{\omega_{\tiny{\textrm{ex}}}\omega_{\tiny{\textrm{A}}}}{2}(\mathbf{l}\cdot\mathbf{z})^2+\frac{\gamma^2}{2}(\mathbf{H_{\tiny{\textrm{DMI}}}}\cdot\mathbf{l})^2\\
    +\frac{\omega_{\tiny{\textrm{ex}}}\omega_{\tiny{\textrm{me}}}}{2}(\mathbf{l}\cdot\mathbf{x})^2+\frac{\gamma^2}{2}(\mathbf{H_0}\cdot\mathbf{l})^2.
    \end{array}
\end{equation}

Here characteristic frequencies are $\omega_{\tiny{\textrm{ex}}}=2\gamma \textrm{H}_{\tiny{\textrm{ex}}}$, $\omega_{\tiny{\textrm{A}}}=\gamma \textrm{H}_{\tiny{\textrm{A}}}$, $\omega_{\tiny{\textrm{me}}}=\gamma \textrm{H}_{\tiny{\textrm{me}}}$, and $\textrm{H}_{\tiny{\textrm{ex}}}$ is the AFM internal exchange magnetic field, $\mathbf{H_{\tiny{\textrm{A}}}}=\textrm{H}_{\tiny{\textrm{A}}}\mathbf{y}$ is the AFM anisotropy field, $\mathbf{H_{\tiny{\textrm{me}}}}=\textrm{H}_{\tiny{\textrm{me}}}\mathbf{x}$ is the effective magnetostriction field\,\cite{Morrish1995}. For the weak ferromagnets, such as hematite, iron borate, manganese carbonate, etc., having the space group of symmetry $D_{3d}^6$, the field of spontaneous magnetostriction, which determines the gap in the spectrum of magnetic excitations at zero field, should be chosen as $\textrm{H}_{\tiny{\textrm{me}}}$. Magnetic moment is always directed along the external magnetic field and the direction of spontaneous striction ``follows'' the direction of the external magnetic field. 

The low-frequency resonant mode of the AFM, leading from the solving of the equation (2) (see for more details Suppl. Inf.) taking into account the applied constant magnetic field and DMI field, is equal to\,\cite{Turov2004}
\begin{equation}\label{eq:04}
\begin{array}{c}
    \omega_{\tiny{\textrm{QFMR}}} = \sqrt{\omega_{\tiny{\textrm{H}}}^2+\omega_{\tiny{\textrm{H}}}\omega_{\tiny{\textrm{DMI}}}+\omega_{\tiny{\textrm{ex}}}\omega_{\tiny{\textrm{me}}}}.
    \end{array}
    \end{equation}

\begin{figure*}[t]  
\includegraphics[width=1\linewidth]{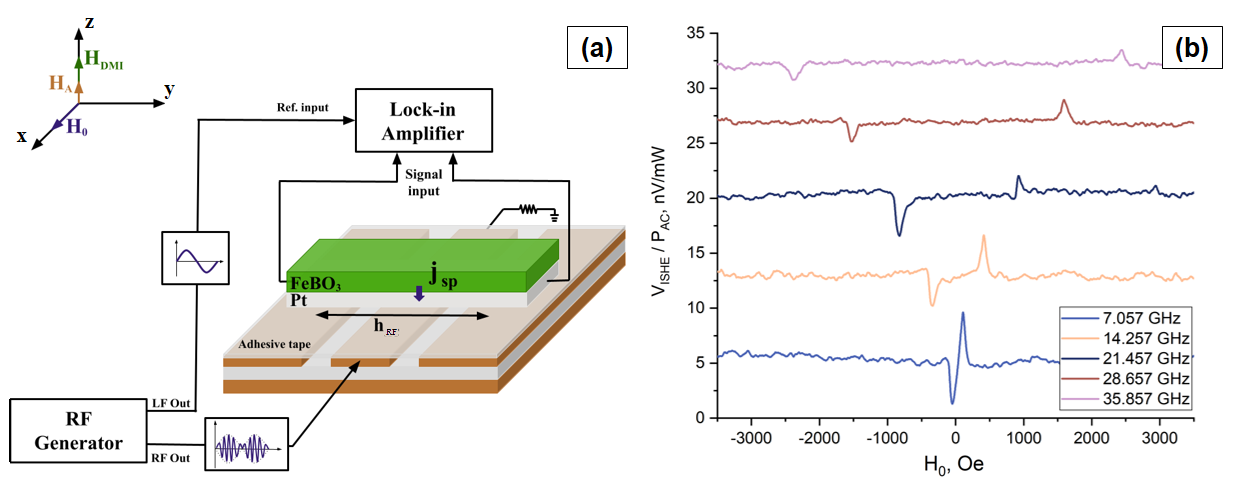}
\caption{
(a) Experimental setup for measuring the ISHE voltage by phase-locking method. A RF signal generator up to 43.5 GHz is used as a RF signal source. The amplitude-modulated signal is applied to the coplanar waveguide. The output of the amplitude modulation signal of the generator is connected to the lock-in amplifier (LIA) reference input thus LIA is phase-locked at the generator modulation frequency. Under the action of the amplitude-modulated signal, a homogeneous precession of the magnetic sublattices of FeBO$_3$ is initiated, which leads to spin-pumping at the boundary of FeBO$_3$/Pt and the emergence of a spin current $\mathbf{j}_{\tiny{\textrm{sp}}}$ in the platinum layer. Due to the inverse spin Hall effect, the appearance of spin current in Pt leads to the emergence of ISHE voltage. (b) Experimentally obtained ISHE voltage of FeBO$_3$/Pt at different frequencies normalized by actual RF power with offset voltage.
}
\label{fig:002}
\end{figure*}

It should be noted that the frequency of quasi-ferromagnetic resonance is significantly affected by the DMI field, which is mostly present in the canted antiferromagnets, as well as the magnetoelastic coupling. The iron borate has a large value of the latter. The application of external mechanical strain allows a significant increase in the resonant frequency.

In order to compare the obtained theoretical expressions with the experimantal results, we measure the ferromagnetic resonance (FMR) using the \textit{VNA-FMR} method (see Suppl. Inf. for more details). We use two port Vector Network Analyser (VNA) up to 20 GHz for radio frequency (RF) signal generation and detection. VNA’s ports are connected to the coplanar waveguide designed for frequencies up to 50 GHz, which in turn is located between the electromagnet’s poles. We measure real and imaginary parts of S$_{21}$-parameter at the fix frequency with magnetic field sweeping (see more details in Supplemental material). Also, the influence of irregularities in the amplitude-frequency response of the microwave path is taken into account. Fig.\,\ref{fig:01}(a) shows the theoretical and experimental dependence of the resonant frequency of FeBO$_3$ for the low-frequency mode on the applied external magnetic field. These experimental points are derived from the FMR spectra (Fig.\,\ref{fig:01}(b)) of FeBO$_3$, measured by the \textit{VNA-FMR} method. As it is shown in Fig.\,\ref{fig:01}(a) theoretical and experimental results are consistent with high accuracy. From FMR spectra we can determine the linewidth. In our case, the linewidth is 50 Oe at 10 GHz, which is significantly smaller than in a similar experiment with hematite \cite{Wang2021,Boventer2021,Lebrun2020}. It is also worth noting a slight increase in the linewidth with a rising frequency of input signal.

Let us calculate the inverse spin Hall voltage caused by the spin pumping from the AFM. As mentioned earlier, an external variable microwave wave with linear polarization initiates a homogeneous precession of the magnetic sublattices $\mathbf{M}_{1,2}$ in the AFM. This precession causes spin pumping at the $\textrm{FeBO}_3/\textrm{Pt}$ boundary and the excitation of the spin current $\mathbf{j}_\textrm{sp}$ in the $\textrm{Pt}$ layer \cite{Baltz2018}:
\begin{equation}\label{eq:05}
\begin{array}{c}
    \mathbf{j}_{\tiny{\textrm{sp}}}=\frac{h g_r}{4 \pi^2}\left[\frac{d\mathbf{l}}{dt}\times\mathbf{l}+\frac{d\mathbf{m}}{dt}\times\mathbf{m}\right],
    \end{array}
\end{equation}
where $g_r$ is the real part of the spin-mixing conductance and $h$ is the Planck constant. Due to the inverse spin Hall effect (ISHE), the appearance of spin current in Pt leads to a non-zero potential difference at the output electrodes $V_{\text{out}}$ placed on the distance $L$. Note that the main contribution to the spin current will be made by the second term on the right side of expression (5), that is, the influence due to a varying of the magnetic moment $\mathbf{m}(t)$, in contrast to most experiments with spin pumping from antiferromagnets that do not have weak ferromagnetism\,\cite{Vaidya2020,Ross2015,Stremoukhov2022}. As can be seen, expression (5) determines the quadratic nature of the spin current dependence on the magnitude of the resonant amplitude of the magnetic moment. In (5) we do not consider the imaginary part of the spin mixing conductance, the corresponding term is proportional to the first derivative of magnetization and does not affect the DC component of the rectified voltage\,\cite{Baltz2018}.

We experimentally demonstrate the phenomenon described above using an installation illustrated in Fig.\,\ref{fig:002}(a). For this measurement we utilize the lock-in technique. A RF signal generator up to 43.5 GHz is used as the RF signal source. The amplitude-modulated signal is applied to the coplanar waveguide, the board of which has two contacts for ISHE voltage measurement. Output of the amplitude modulation signal of the generator is connected to the lock-in amplifier (LIA) reference input thus LIA is phase-locked on the modulation frequency of the generator. The sample was separated from the coplanar waveguide by adhesive tape due to the conductivity of platinum layer.

\begin{figure*}[t]  
\includegraphics[width=1\linewidth]{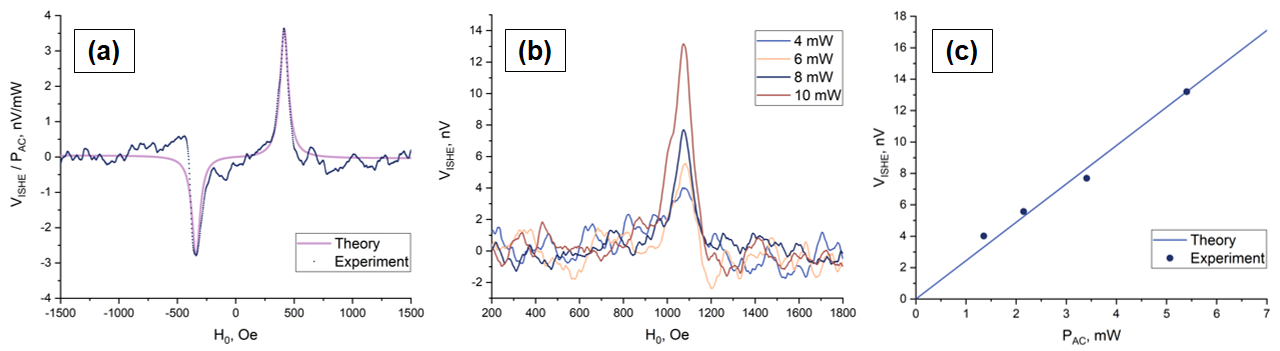}
\caption{
(a) Comparison of the theoretical curve obtained by Eq. (8) with the experimental points of ISHE voltage at 14.257 GHz, normalized by actual RF power. (b) Experimentally measured dependence of ISHE voltage on the applied external magnetic field at various RF power at 23.514 GHz. (c) Theoretical and experimentally measured power dependence of the ISHE voltage at 23.514 GHz.
}
\label{fig:03}
\end{figure*}

We obtain a set of experimentally received dependencies of ISHE voltage on different frequencies normalized by actual RF power with offset voltage from magnetic field (Fig.\,\ref{fig:002}(b)). One can see, as the frequency increases, the resonant peaks drift apart, and their amplitude decreases. An important feature is that at positive magnetic field values, the resonant peak is also positive, and at negative magnetic fields, it is negative.

For a more detailed study of this behavior let us derive an expression for the inverse spin Hall voltage $V_{\tiny{\textrm{out}}}$ induced by spin pumping from the AFM into the HM layer \cite{Safin2022,Cheng2014,Cheng2016}:
\begin{equation}\label{eq:07}
\begin{array}{c}
    V_{\tiny{\textrm{out}}}=\left.\kappa\left<\frac{d\mathbf{l}}{dt}\times\mathbf{l}+\frac{d\mathbf{m}}{dt}\times\mathbf{m}\right>\right|_{\mathbf{x}},
    \end{array}
    \end{equation}
where $\left<...\right>$ is average of time during the period $2\pi/\omega_{\tiny{\textrm{QFMR}}}$, $\kappa$ is the proportionality coefficient\,\cite{Khymyn2017}:
\begin{equation}\label{eq:08}
\begin{array}{c}
    \kappa = \frac{L g_r \theta_{\tiny{\textrm{SH}}} e \lambda_{\tiny{\textrm{Pt}}} \rho}{2 \pi d_{\tiny{\textrm{Pt}}}} \tanh\left(\frac{d_{\tiny{\textrm{Pt}}}}{2 \lambda_{\tiny{\textrm{Pt}}}}\right),
    \end{array}
    \end{equation}
and $L$ is the distance between output electrodes, $\theta_{\tiny{\textrm{SH}}}$ is the spin-Hall angle, $e$ is the electron charge, $\lambda_{\tiny{\textrm{Pt}}}$ is the spin-diffusion length, while $\rho$ and $d_{\tiny{\textrm{Pt}}}$ are the electrical resistivity and thickness of the Pt layer, respectively.

For the input alternating power of the RF signal $P_{\tiny{\textrm{RF}}}=\frac{c}{2\mu_0}S\cdot \left(h_{\tiny{\textrm{RF}}}\right)^2$, where $c$ is the speed of light, $\mu_0$ is the vacuum magnetic permeability, $S$ is the AFM layer cross-section, one can find the detector sensitivity defined as (see Suppl. Inf.):
\begin{equation}\label{eq:09}
\begin{array}{c}
    R\left(\omega\right) = \frac{\left|V_{\tiny{\textrm{out}}}\left(\omega\right)\right|}{P_{\tiny{\textrm{RF}}}}= \frac{2\mu_0\gamma^2\kappa\cdot\kappa^{'}}{cS \omega_{\tiny{\textrm{ex}}}^2}\cdot\frac{\left(\omega_{\tiny{\textrm{H}}}+\omega_{\tiny{\textrm{DMI}}}\right)\omega_{\tiny{\textrm{DMI}}}^2}{\left(\omega^2-\omega_{\tiny{\textrm{QFMR}}}^2\right)^2+\left(\gamma_{\tiny{\textrm{eff}}}\omega\right)^2}\omega^2.
    \end{array}
    \end{equation}

\noindent where $\kappa^{'}$ is phenomenological fit parameter that has been obtained by comparing experimental data with theoretical model. By analyzing the expression, we can determine that the sensitivity's dependence on the magnetic field is a linear function. All numerical values of constants used for constructing theoretical curves are given in Table I\,\cite{LVVelikov1974,Popov2015}. Some parameters, i.e., the spin Hall $\theta_{SH}$ angle and the spin mixing constant $g_{r}$, were taken to coincide with similar parameters for hematite\,\cite{Boventer2021,Wang2021}.
Fig.\,\ref{fig:03}(a) shows a comparison of the theoretical curve obtained by Eq. (8) with the experimental dependence of ISHE voltage on the applied external magnetic field at a frequency of 14.257 GHz. A number of curves is measured (Fig.\,\ref{fig:03}(b)) at different RF power to determine the nature of the change in the ISHE voltage level from the applied RF power (Fig.\,\ref{fig:03}(c)) at 23.514 GHz.
\newline
\newline
\noindent TABLE I. Numerical values of the parameters used to calculate the characteristics for the FeBO$_3$\,\cite{LVVelikov1974,Popov2015,Boventer2021,Wang2021}.
\newline
\newline 
\begin{tabular}{ |p{1.9 cm}|p{2.2cm}|p{4cm}| }
 \hline
 \textbf{Parameter} & \textbf{Value} & \textbf{Physical meaning} \\
 \hline
 $\theta_\textrm{SH}$  & 0.1  & Spin-Hall angle  \\
 \hline
 $\textrm{g}_\textrm{r}$  & 6 $\cdot 10^{18}$ m$^{-2}$  & Real part of the spin-mixing conductance  \\
 \hline
 $\alpha_\textrm{eff}$  & $5.5\cdot10^{-3}$  & Effective damping including Gilbert constant and spin-pumping terms  \\
 \hline
 $\textrm{H}_\textrm{ex}$  & 6500 kOe  & Exchange magnetic field  \\
 \hline
 $\textrm{H}_\textrm{me}$  & 0.2 Oe  & Effective magnetostriction field  \\
 \hline
  $\textrm{H}_\textrm{DMI}$  & 100 kOe  & DMI field  \\
 \hline
 $\lambda_\textrm{Pt}$  & 1.2 nm  & Spin-diffusion length  \\
 \hline
 $\textrm{d}_\textrm{Pt}$  & 2 nm  & Thickness of the Pt layer  \\
 \hline
 $\rho$  & $2.5 \cdot 10^{-7} \Omega \cdot$ m  & Electrical resistivity of the Pt layer  \\
 \hline
 $L$  & 3 mm  & Distance between output electrodes  \\
 \hline
 S  & 5 $\times$ 5  mm$^2$  & AFM layer cross-section  \\
\hline
\end{tabular}
\newline

Comparison of theoretical and experimental results presented in Fig.\,\ref{fig:03}(a) shows a good confirmation of the theoretical model presented in current article. Small differences are due to the parasitic effect of the output contacts in the experimental measurement, which is not taken into account in the theoretical description. Eq. (8) shows that when the input frequency increases, a larger magnetic field is required to achieve resonance, which in turn leads to a decrease in the resonance peak amplitude. Fig.\,\ref{fig:03}(b) demonstrates the experimentally measured dependence of ISHE voltage on the applied external magnetic field at various RF power. This dependence has a resonant nature. Also, it can be noticed that the ISHE voltage amplitude and the linewidth are grow by increasing the power of the RF signal. The theoretical and experimental dependence of ISHE voltage on RF signal power can be seen in Fig.\,\ref{fig:03}(c). The slope of this characteristic determines the volt-watt sensitivity of the structure. ISHE voltage at a frequency of 23.514 GHz is 13.2 nV, which at an input power of 5.403 mW corresponds to the sensitivity of 2.5 $\mu$V/W. 

Finally, we have demonstrated both theoretically and experimentally that an antiferromagnet with weak ferromagnetism FeBO$_3$ can be used as a sensitive element to detect the resonance of a linearly polarized RF wave in tens of GHz. We studied FMR spectra of this structure and showed that the theoretical and experimental dependence of resonance frequency from magnetic field were consistent. In addition, we measured ISHE voltage from the structure and compared it to the theoretical model. It was noted that an increase in the frequency of input microwave field causes an increase in the distance between the resonance peaks as well as a decrease in their amplitude. A study of the ISHE voltage dependence on the magnetic field showed that an increase in the power of the applied RF signal increases the amplitude of the resonance peak as well as the width of the line. We obtained a value of the sensitivity of 2.5 $\mu$V/W, typical of similar structures (for example, in \cite{Boventer2021} a value of the sensitivity is about 0.4 $\mu$V/W). Comparing our results with the results of similar works, we can conclude that they are consistent with each other, but we demonstrated a narrower linewidth. In our opinion, the study of structures based on antiferromagnets with weak ferromagnetism allows to expand the understanding of spin dynamics, as well as to implement various microwave and terahertz applications, developments of antiferromagnetic spintronics, magnonics, and data storage.

The work was carried out within the framework of the state assignment of the Kotel'nikov Institute of Radio Engineering and Electronics of the Russian Academy of Sciences

\bibliography{main}

\providecommand{\noopsort}[1]{}\providecommand{\singleletter}[1]{#1}%
\begin{thebibliography}{56}
\providecommand{\natexlab}[1]{#1}
\providecommand{\url}[1]{\texttt{#1}}
\expandafter\ifx\csname urlstyle\endcsname\relax
  \providecommand{\doi}[1]{doi: #1}\else
  \providecommand{\doi}{doi: \begingroup \urlstyle{rm}\Url}\fi

\bibitem[Saitoh et~al.(2006)Saitoh, Ueda, Miyajima, and Tatara]{Saitoh2006}
E.~Saitoh, M.~Ueda, H.~Miyajima, and G.~Tatara.
\newblock Conversion of spin current into charge current at room temperature:
  Inverse spin-hall effect.
\newblock \emph{Applied Physics Letters}, 88\penalty0 (18), May 2006.
\newblock \doi{10.1063/1.2199473}.
\newblock URL \url{https://doi.org/10.1063/1.2199473}.

\bibitem[Azevedo et~al.(2005)Azevedo, Le{\~{a}}o, Rodriguez-Suarez, Oliveira,
  and Rezende]{Azevedo2005}
A.~Azevedo, L.~H.~Vilela Le{\~{a}}o, R.~L. Rodriguez-Suarez, A.~B. Oliveira,
  and S.~M. Rezende.
\newblock dc effect in ferromagnetic resonance: Evidence of the spin-pumping
  effect?
\newblock \emph{Journal of Applied Physics}, 97\penalty0 (10), May 2005.
\newblock \doi{10.1063/1.1855251}.
\newblock URL \url{https://doi.org/10.1063/1.1855251}.

\bibitem[Heinrich et~al.(2011)Heinrich, Burrowes, Montoya, Kardasz, Girt, Song,
  Sun, and Wu]{Heinrich2011}
B.~Heinrich, C.~Burrowes, E.~Montoya, B.~Kardasz, E.~Girt, Young-Yeal Song,
  Yiyan Sun, and Mingzhong Wu.
\newblock Spin pumping at the magnetic insulator ({YIG})/normal metal (au)
  interfaces.
\newblock \emph{Physical Review Letters}, 107\penalty0 (6), August 2011.
\newblock \doi{10.1103/physrevlett.107.066604}.
\newblock URL \url{https://doi.org/10.1103/physrevlett.107.066604}.

\bibitem[Takahashi(2016)]{Takahashi2016}
Saburo Takahashi.
\newblock Physical principles of spin pumping.
\newblock In \emph{Handbook of Spintronics}, pages 1445--1480. Springer
  Netherlands, 2016.
\newblock \doi{10.1007/978-94-007-6892-5_51}.
\newblock URL \url{https://doi.org/10.1007/978-94-007-6892-5_51}.

\bibitem[Xiong et~al.(2022)Xiong, Jiang, Shi, Du, Yao, Guo, Zhu, Cao, Peng,
  Cai, Zhu, and Zhao]{Xiong2022}
Danrong Xiong, Yuhao Jiang, Kewen Shi, Ao~Du, Yuxuan Yao, Zongxia Guo, Daoqian
  Zhu, Kaihua Cao, Shouzhong Peng, Wenlong Cai, Dapeng Zhu, and Weisheng Zhao.
\newblock Antiferromagnetic spintronics: An overview and outlook.
\newblock \emph{Fundamental Research}, 2\penalty0 (4):\penalty0 522--534, July
  2022.
\newblock \doi{10.1016/j.fmre.2022.03.016}.
\newblock URL \url{https://doi.org/10.1016/j.fmre.2022.03.016}.

\bibitem[Baltz et~al.(2018)Baltz, Manchon, Tsoi, Moriyama, Ono, and
  Tserkovnyak]{Baltz2018}
V.~Baltz, A.~Manchon, M.~Tsoi, T.~Moriyama, T.~Ono, and Y.~Tserkovnyak.
\newblock Antiferromagnetic spintronics.
\newblock \emph{Reviews of Modern Physics}, 90\penalty0 (1), February 2018.
\newblock \doi{10.1103/revmodphys.90.015005}.
\newblock URL \url{https://doi.org/10.1103/revmodphys.90.015005}.

\bibitem[Jungwirth et~al.(2016)Jungwirth, Marti, Wadley, and
  Wunderlich]{Jungwirth2016}
T.~Jungwirth, X.~Marti, P.~Wadley, and J.~Wunderlich.
\newblock Antiferromagnetic spintronics.
\newblock \emph{Nature Nanotechnology}, 11\penalty0 (3):\penalty0 231--241,
  March 2016.
\newblock \doi{10.1038/nnano.2016.18}.
\newblock URL \url{https://doi.org/10.1038/nnano.2016.18}.

\bibitem[Rezende et~al.(2019)Rezende, Azevedo, and
  Rodr{\'{\i}}guez-Su{\'{a}}rez]{Rezende2019}
Sergio~M. Rezende, Antonio Azevedo, and Roberto~L.
  Rodr{\'{\i}}guez-Su{\'{a}}rez.
\newblock Introduction to antiferromagnetic magnons.
\newblock \emph{Journal of Applied Physics}, 126\penalty0 (15), October 2019.
\newblock \doi{10.1063/1.5109132}.
\newblock URL \url{https://doi.org/10.1063/1.5109132}.

\bibitem[Barman et~al.(2021)Barman, Gubbiotti, Ladak, Adeyeye, Krawczyk,
  Gr\"{a}fe, Adelmann, Cotofana, Naeemi, Vasyuchka, Hillebrands, Nikitov, Yu,
  Grundler, Sadovnikov, Grachev, Sheshukova, Duquesne, Marangolo, Csaba, Porod,
  Demidov, Urazhdin, Demokritov, Albisetti, Petti, Bertacco, Schultheiss,
  Kruglyak, Poimanov, Sahoo, Sinha, Yang, M\"{u}nzenberg, Moriyama, Mizukami,
  Landeros, Gallardo, Carlotti, Kim, Stamps, Camley, Rana, Otani, Yu, Yu,
  Bauer, Back, Uhrig, Dobrovolskiy, Budinska, Qin, van Dijken, Chumak, Khitun,
  Nikonov, Young, Zingsem, and Winklhofer]{Barman2021}
Anjan Barman, Gianluca Gubbiotti, S~Ladak, A~O Adeyeye, M~Krawczyk,
  J~Gr\"{a}fe, C~Adelmann, S~Cotofana, A~Naeemi, V~I Vasyuchka, B~Hillebrands,
  S~A Nikitov, H~Yu, D~Grundler, A~V Sadovnikov, A~A Grachev, S~E Sheshukova,
  J-Y Duquesne, M~Marangolo, G~Csaba, W~Porod, V~E Demidov, S~Urazhdin, S~O
  Demokritov, E~Albisetti, D~Petti, R~Bertacco, H~Schultheiss, V~V Kruglyak,
  V~D Poimanov, S~Sahoo, J~Sinha, H~Yang, M~M\"{u}nzenberg, T~Moriyama,
  S~Mizukami, P~Landeros, R~A Gallardo, G~Carlotti, J-V Kim, R~L Stamps, R~E
  Camley, B~Rana, Y~Otani, W~Yu, T~Yu, G~E~W Bauer, C~Back, G~S Uhrig, O~V
  Dobrovolskiy, B~Budinska, H~Qin, S~van Dijken, A~V Chumak, A~Khitun, D~E
  Nikonov, I~A Young, B~W Zingsem, and M~Winklhofer.
\newblock The 2021 magnonics roadmap.
\newblock \emph{Journal of Physics: Condensed Matter}, 33\penalty0
  (41):\penalty0 413001, August 2021.
\newblock \doi{10.1088/1361-648x/abec1a}.
\newblock URL \url{https://doi.org/10.1088/1361-648x/abec1a}.

\bibitem[Cheng et~al.(2016)Cheng, Xiao, and Brataas]{Cheng2016}
Ran Cheng, Di~Xiao, and Arne Brataas.
\newblock Terahertz antiferromagnetic spin hall nano-oscillator.
\newblock \emph{Physical Review Letters}, 116\penalty0 (20), May 2016.
\newblock \doi{10.1103/physrevlett.116.207603}.
\newblock URL \url{https://doi.org/10.1103/physrevlett.116.207603}.

\bibitem[Khymyn et~al.(2017{\natexlab{a}})Khymyn, Lisenkov, Tiberkevich,
  Ivanov, and Slavin]{RomanKhymyn2017}
Roman Khymyn, Ivan Lisenkov, Vasyl Tiberkevich, Boris~A. Ivanov, and Andrei
  Slavin.
\newblock Antiferromagnetic {THz}-frequency josephson-like oscillator driven by
  spin current.
\newblock \emph{Scientific Reports}, 7\penalty0 (1), March 2017{\natexlab{a}}.
\newblock \doi{10.1038/srep43705}.
\newblock URL \url{https://doi.org/10.1038/srep43705}.

\bibitem[Sulymenko et~al.(2017)Sulymenko, Prokopenko, Tiberkevich, Slavin,
  Ivanov, and Khymyn]{Sulymenko2017}
O.R. Sulymenko, O.V. Prokopenko, V.S. Tiberkevich, A.N. Slavin, B.A. Ivanov,
  and R.S. Khymyn.
\newblock Terahertz-frequency spin hall auto-oscillator based on a canted
  antiferromagnet.
\newblock \emph{Physical Review Applied}, 8\penalty0 (6), December 2017.
\newblock \doi{10.1103/physrevapplied.8.064007}.
\newblock URL \url{https://doi.org/10.1103/physrevapplied.8.064007}.

\bibitem[Popov et~al.(2020)Popov, Safin, Kirilyuk, Nikitov, Lisenkov,
  Tyberkevich, and Slavin]{Popov2020}
P.A. Popov, A.R. Safin, A.~Kirilyuk, S.A. Nikitov, I.~Lisenkov, V.~Tyberkevich,
  and A.~Slavin.
\newblock Voltage-controlled anisotropy and current-induced magnetization
  dynamics in antiferromagnetic-piezoelectric layered heterostructures.
\newblock \emph{Physical Review Applied}, 13\penalty0 (4), April 2020.
\newblock \doi{10.1103/physrevapplied.13.044080}.
\newblock URL \url{https://doi.org/10.1103/physrevapplied.13.044080}.

\bibitem[Consolo et~al.(2021)Consolo, Valenti, Safin, Nikitov, Tyberkevich, and
  Slavin]{Consolo2021}
G.~Consolo, G.~Valenti, A.~R. Safin, S.~A. Nikitov, V.~Tyberkevich, and
  A.~Slavin.
\newblock Theory of the electric field controlled antiferromagnetic spin hall
  oscillator and detector.
\newblock \emph{Physical Review B}, 103\penalty0 (13), April 2021.
\newblock \doi{10.1103/physrevb.103.134431}.
\newblock URL \url{https://doi.org/10.1103/physrevb.103.134431}.

\bibitem[Khymyn et~al.(2017{\natexlab{b}})Khymyn, Tiberkevich, and
  Slavin]{Khymyn2017}
Roman Khymyn, Vasil Tiberkevich, and Andrei Slavin.
\newblock Antiferromagnetic spin current rectifier.
\newblock \emph{{AIP} Advances}, 7\penalty0 (5), March 2017{\natexlab{b}}.
\newblock \doi{10.1063/1.4977974}.
\newblock URL \url{https://doi.org/10.1063/1.4977974}.

\bibitem[Gomonay et~al.(2018)Gomonay, Jungwirth, and Sinova]{Gomonay2018}
O.~Gomonay, T.~Jungwirth, and J.~Sinova.
\newblock Narrow-band tunable terahertz detector in antiferromagnets via
  staggered-field and antidamping torques.
\newblock \emph{Physical Review B}, 98\penalty0 (10), September 2018.
\newblock \doi{10.1103/physrevb.98.104430}.
\newblock URL \url{https://doi.org/10.1103/physrevb.98.104430}.

\bibitem[Safin et~al.(2020)Safin, Puliafito, Carpentieri, Finocchio, Nikitov,
  Stremoukhov, Kirilyuk, Tyberkevych, and Slavin]{Safin2020}
A.~Safin, V.~Puliafito, M.~Carpentieri, G.~Finocchio, S.~Nikitov,
  P.~Stremoukhov, A.~Kirilyuk, V.~Tyberkevych, and A.~Slavin.
\newblock Electrically tunable detector of {THz}-frequency signals based on an
  antiferromagnet.
\newblock \emph{Applied Physics Letters}, 117\penalty0 (22), November 2020.
\newblock \doi{10.1063/5.0031053}.
\newblock URL \url{https://doi.org/10.1063/5.0031053}.

\bibitem[Safin et~al.(2022)Safin, Nikitov, Kirilyuk, Tyberkevych, and
  Slavin]{Safin2022}
Ansar Safin, Sergey Nikitov, Andrei Kirilyuk, Vasyl Tyberkevych, and Andrei
  Slavin.
\newblock Theory of antiferromagnet-based detector of terahertz frequency
  signals.
\newblock \emph{Magnetochemistry}, 8\penalty0 (2):\penalty0 26, February 2022.
\newblock \doi{10.3390/magnetochemistry8020026}.
\newblock URL \url{https://doi.org/10.3390/magnetochemistry8020026}.

\bibitem[Kozlova et~al.(2022)Kozlova, Safin, and Nikitov]{Kozlova2022}
E.~E. Kozlova, A.~R. Safin, and S.~A. Nikitov.
\newblock Ferrimagnet based spin hall detector of subterahertz frequency
  signals.
\newblock \emph{Applied Physics Letters}, 121\penalty0 (19), November 2022.
\newblock \doi{10.1063/5.0112050}.
\newblock URL \url{https://doi.org/10.1063/5.0112050}.

\bibitem[Stremoukhov et~al.(2019)Stremoukhov, Safin, Logunov, Nikitov, and
  Kirilyuk]{Stremoukhov2019}
P.~Stremoukhov, A.~Safin, M.~Logunov, S.~Nikitov, and A.~Kirilyuk.
\newblock Spintronic terahertz-frequency nonlinear emitter based on the canted
  antiferromagnet-platinum bilayers.
\newblock \emph{Journal of Applied Physics}, 125\penalty0 (22), June 2019.
\newblock \doi{10.1063/1.5090455}.
\newblock URL \url{https://doi.org/10.1063/1.5090455}.

\bibitem[Khymyn et~al.(2016)Khymyn, Lisenkov, Tiberkevich, Slavin, and
  Ivanov]{Khymyn2016}
Roman Khymyn, Ivan Lisenkov, Vasil~S. Tiberkevich, Andrei~N. Slavin, and
  Boris~A. Ivanov.
\newblock Transformation of spin current by antiferromagnetic insulators.
\newblock \emph{Physical Review B}, 93\penalty0 (22), June 2016.
\newblock \doi{10.1103/physrevb.93.224421}.
\newblock URL \url{https://doi.org/10.1103/physrevb.93.224421}.

\bibitem[Bradley et~al.(2023)Bradley, Louis, Trevillian, Quach, Bankowski,
  Slavin, and Tyberkevych]{Bradley2023}
H.~Bradley, S.~Louis, C.~Trevillian, L.~Quach, E.~Bankowski, A.~Slavin, and
  V.~Tyberkevych.
\newblock Artificial neurons based on antiferromagnetic auto-oscillators as a
  platform for neuromorphic computing.
\newblock \emph{{AIP} Advances}, 13\penalty0 (1), January 2023.
\newblock \doi{10.1063/5.0128530}.
\newblock URL \url{https://doi.org/10.1063/5.0128530}.

\bibitem[Sulymenko et~al.(2018)Sulymenko, Prokopenko, Lisenkov, {\AA}kerman,
  Tyberkevych, Slavin, and Khymyn]{Sulymenko2018}
Olga Sulymenko, Oleksandr Prokopenko, Ivan Lisenkov, Johan {\AA}kerman, Vasyl
  Tyberkevych, Andrei~N. Slavin, and Roman Khymyn.
\newblock Ultra-fast logic devices using artificial neurons based on
  antiferromagnetic pulse generators.
\newblock \emph{Journal of Applied Physics}, 124\penalty0 (15), September 2018.
\newblock \doi{10.1063/1.5042348}.
\newblock URL \url{https://doi.org/10.1063/1.5042348}.

\bibitem[Mitrofanova et~al.(2022)Mitrofanova, Safin, Kravchenko, Nikitov, and
  Kirilyuk]{Mitrofanova2022}
A.~Mitrofanova, A.~Safin, O.~Kravchenko, S.~Nikitov, and A.~Kirilyuk.
\newblock Optically initialized and current-controlled logical element based on
  antiferromagnetic-heavy metal heterostructures for neuromorphic computing.
\newblock \emph{Applied Physics Letters}, 120\penalty0 (7), February 2022.
\newblock \doi{10.1063/5.0079532}.
\newblock URL \url{https://doi.org/10.1063/5.0079532}.

\bibitem[Kosub et~al.(2017)Kosub, Kopte, H\"{u}hne, Appel, Shields, Maletinsky,
  H\"{u}bner, Liedke, Fassbender, Schmidt, and Makarov]{Kosub2017}
Tobias Kosub, Martin Kopte, Ruben H\"{u}hne, Patrick Appel, Brendan Shields,
  Patrick Maletinsky, Ren{\'{e}} H\"{u}bner, Maciej~Oskar Liedke, J\"{u}rgen
  Fassbender, Oliver~G. Schmidt, and Denys Makarov.
\newblock Purely antiferromagnetic magnetoelectric random access memory.
\newblock \emph{Nature Communications}, 8\penalty0 (1), January 2017.
\newblock \doi{10.1038/ncomms13985}.
\newblock URL \url{https://doi.org/10.1038/ncomms13985}.

\bibitem[Fina et~al.(2020)Fina, Dix, Men{\'{e}}ndez, Crespi, Foerster, Aballe,
  S{\'{a}}nchez, and Fontcuberta]{Fina2020}
Ignasi Fina, Nico Dix, Enric Men{\'{e}}ndez, Anna Crespi, Michael Foerster,
  Lucia Aballe, Florencio S{\'{a}}nchez, and Josep Fontcuberta.
\newblock Flexible antiferromagnetic {FeRh} tapes as memory elements.
\newblock \emph{{ACS} Applied Materials $\&$ Interfaces}, 12\penalty0
  (13):\penalty0 15389--15395, March 2020.
\newblock \doi{10.1021/acsami.0c00704}.
\newblock URL \url{https://doi.org/10.1021/acsami.0c00704}.

\bibitem[Artemchuk et~al.(2020)Artemchuk, Sulymenko, Louis, Li, Khymyn,
  Bankowski, Meitzler, Tyberkevych, Slavin, and Prokopenko]{Artemchuk2020}
P.~Yu. Artemchuk, O.~R. Sulymenko, S.~Louis, J.~Li, R.~S. Khymyn, E.~Bankowski,
  T.~Meitzler, V.~S. Tyberkevych, A.~N. Slavin, and O.~V. Prokopenko.
\newblock Terahertz frequency spectrum analysis with a nanoscale
  antiferromagnetic tunnel junction.
\newblock \emph{Journal of Applied Physics}, 127\penalty0 (6), February 2020.
\newblock URL \url{https://doi.org/10.1063/1.5140552}.

\bibitem[Mitrofanova et~al.(2023)Mitrofanova, Vanin, Volkov, Safin, Kravchenko,
  Ryu, and Nikitov]{Mitrofanova2023}
Anastasia Mitrofanova, Kirill Vanin, Dmitry Volkov, Ansar Safin, Oleg
  Kravchenko, Heung-Gyoon Ryu, and Sergey Nikitov.
\newblock Spectral analysis of subterahertz oscillations based on an
  antiferromagnet/non{\textendash}magnetic metal heterostructure.
\newblock \emph{{IEEE} Transactions on Nanotechnology}, 22:\penalty0 620--627,
  2023.
\newblock URL \url{https://doi.org/10.1109/tnano.2023.3315071}.

\bibitem[Li et~al.(2020)Li, Wilson, Cheng, Lohmann, Kavand, Yuan, Aldosary,
  Agladze, Wei, Sherwin, and Shi]{Li2020}
Junxue Li, C.~Blake Wilson, Ran Cheng, Mark Lohmann, Marzieh Kavand, Wei Yuan,
  Mohammed Aldosary, Nikolay Agladze, Peng Wei, Mark~S. Sherwin, and Jing Shi.
\newblock Spin current from sub-terahertz-generated antiferromagnetic magnons.
\newblock \emph{Nature}, 578\penalty0 (7793):\penalty0 70--74, January 2020.
\newblock \doi{10.1038/s41586-020-1950-4}.
\newblock URL \url{https://doi.org/10.1038/s41586-020-1950-4}.

\bibitem[Vaidya et~al.(2020)Vaidya, Morley, van Tol, Liu, Cheng, Brataas,
  Lederman, and del Barco]{Vaidya2020}
Priyanka Vaidya, Sophie~A. Morley, Johan van Tol, Yan Liu, Ran Cheng, Arne
  Brataas, David Lederman, and Enrique del Barco.
\newblock Subterahertz spin pumping from an insulating antiferromagnet.
\newblock \emph{Science}, 368\penalty0 (6487):\penalty0 160--165, April 2020.
\newblock \doi{10.1126/science.aaz4247}.
\newblock URL \url{https://doi.org/10.1126/science.aaz4247}.

\bibitem[Ross et~al.(2015)Ross, Schreier, Lotze, Huebl, Gross, and
  Goennenwein]{Ross2015}
Philipp Ross, Michael Schreier, Johannes Lotze, Hans Huebl, Rudolf Gross, and
  Sebastian T.~B. Goennenwein.
\newblock Antiferromagentic resonance detected by direct current voltages in
  {MnF}2/pt bilayers.
\newblock \emph{Journal of Applied Physics}, 118\penalty0 (23), December 2015.
\newblock \doi{10.1063/1.4937913}.
\newblock URL \url{https://doi.org/10.1063/1.4937913}.

\bibitem[Stremoukhov et~al.(2022)Stremoukhov, Safin, Schippers, Lavrijsen, Bal,
  Zeitler, Sadovnikov, Ilkhchy, Nikitov, and Kirilyuk]{Stremoukhov2022}
P.~Stremoukhov, A.~Safin, C.~F. Schippers, R.~Lavrijsen, M.~Bal, U.~Zeitler,
  A.~Sadovnikov, K.~S. Ilkhchy, S.~Nikitov, and A.~Kirilyuk.
\newblock Strongly nonlinear antiferromagnetic dynamics in high magnetic
  fields.
\newblock \emph{arXiv:2211.00353}, 2022.

\bibitem[Boventer et~al.(2021)Boventer, Simensen, Anane, Klaui, Brataas, and
  Lebrun]{Boventer2021}
I.~Boventer, H.~T. Simensen, A.~Anane, M.~Klaui, A.~Brataas, and R.~Lebrun.
\newblock Room-temperature antiferromagnetic resonance and inverse spin-hall
  voltage in canted antiferromagnets.
\newblock \emph{Phys. Rev. Lett.}, 126:\penalty0 187201, 2021.

\bibitem[Wang et~al.(2021)Wang, Xiao, Guo, Lee-Wong, Yan, Cheng, and
  Du]{Wang2021}
H.~Wang, Y.~Xiao, M.~Guo, E.~Lee-Wong, G.~Q. Yan, R.~Cheng, and C.~R. Du.
\newblock Spin pumping of an easy-plane antiferromagnet enhanced by
  dzyaloshinskii-moriya interaction.
\newblock \emph{Physical Review Letters}, 127, 2021.

\bibitem[Lebrun et~al.(2020)Lebrun, Ross, Gomonay, Baltz, Ebels, Barra,
  Qaiumzadeh, Brataas, Sinova, and Klaui]{Lebrun2020}
R.~Lebrun, A.~Ross, O.~Gomonay, V.~Baltz, U.~Ebels, A.-L. Barra, A.~Qaiumzadeh,
  A.~Brataas, J.~Sinova, and M.~Klaui.
\newblock Long-distance spin-transport across the morin phase transition up to
  room temperature in ultra-low damping single crystals of the antiferromagnet
  $\alpha$-\mbox{Fe$_2$O$_3$}.
\newblock \emph{Nat. Commun.}, 11:\penalty0 6332, 2020.

\bibitem[Turov et~al.(2004)Turov, Kolchanov, Men’shenin, and
  Mirsaev]{Turov2004}
E.~A. Turov, A.~V. Kolchanov, V.~V. Men’shenin, and I.~F. Mirsaev.
\newblock \emph{Symmetry and Physical Properties of Antiferromagnetics}.
\newblock Cambridge International Science Publishing, 2004.

\bibitem[Dmitrienko et~al.(2014)Dmitrienko, Ovchinnikova, Collins, Nisbet,
  Beutier, Kvashnin, Mazurenko, Lichtenstein, and Katsnelson]{Dmitrienko2014}
V.~E. Dmitrienko, E.~N. Ovchinnikova, S.~P. Collins, G.~Nisbet, G.~Beutier,
  Y.~O. Kvashnin, V.~V. Mazurenko, A.~I. Lichtenstein, and M.~I. Katsnelson.
\newblock Measuring the \mbox{Dzyaloshinskii{\textendash}Moriya} interaction in
  a weak ferromagnet.
\newblock \emph{Nature Physics}, 10\penalty0 (3):\penalty0 202--206, 2014.

\bibitem[Kuzmenko et~al.(2014)Kuzmenko, Shuvaev, Dziom, Pimenov, Schiebl,
  Mukhin, Ivanov, Bezmaternykh, and Pimenov]{Kuzmenko2014}
A.~M. Kuzmenko, A.~Shuvaev, V.~Dziom, Anna Pimenov, M.~Schiebl, A.~A. Mukhin,
  V.~Yu. Ivanov, L.~N. Bezmaternykh, and A.~Pimenov.
\newblock Giant gigahertz optical activity in multiferroic ferroborate.
\newblock \emph{Physical Review B}, 89\penalty0 (17):\penalty0 174407, 2014.

\bibitem[Afanasiev et~al.(2014)Afanasiev, Razdolski, Skibinsky, Bolotin,
  Yagupov, Strugatsky, Kirilyuk, Rasing, and Kimel]{Afanasiev2014}
D.~Afanasiev, I.~Razdolski, K.~M. Skibinsky, D.~Bolotin, S.~V. Yagupov, M.~B.
  Strugatsky, A.~Kirilyuk, Th. Rasing, and A.~V. Kimel.
\newblock Laser excitation of lattice-driven anharmonic magnetization dynamics
  in dielectric \mbox{FeBO$_3$}.
\newblock \emph{Physical Review Letters}, 112\penalty0 (14):\penalty0 147403,
  2014.

\bibitem[Velikov et~al.(1974)Velikov, Prokhorov, Rudashevskii, and
  Seleznev]{LVVelikov1974}
L.V. Velikov, A.S. Prokhorov, E.G. Rudashevskii, and V.N. Seleznev.
\newblock Antiferromagnetic resonance in \mbox{FeBO$_3$}.
\newblock \emph{Sov. Phys. JETP}, 39\penalty0 (5), 1974.

\bibitem[Jantz and Wettling(1978)]{Jantz1978}
W.~Jantz and W.~Wettling.
\newblock Spin wave dispersion of \mbox{FeBO$_3$} at small wavevectors.
\newblock \emph{Applied Physics}, 15\penalty0 (4):\penalty0 399--407, 1978.

\bibitem[Diehl(1975)]{Diehl1975}
R.~Diehl.
\newblock Crystal structure refinement of ferric borate, \mbox{FeBO$_3$}.
\newblock \emph{Solid State Communications}, 17\penalty0 (6):\penalty0
  743--745, 1975.

\bibitem[Ovchinnikov et~al.(2020)Ovchinnikov, Rudenko, Kazak, Edelman, and
  Gavrichkov]{Ovchinnikov2020}
S.~G. Ovchinnikov, V.~V. Rudenko, N.~V. Kazak, I.~S. Edelman, and V.~A.
  Gavrichkov.
\newblock Weak antiferromagnet iron borate \mbox{FeBO$_3$}. classic object for
  magnetism and the state of the art.
\newblock \emph{J. Exp. Theor. Phys.}, 131:\penalty0 177--188, 2020.

\bibitem[Petrov et~al.(2021)Petrov, Smolenskii, Paugurt, Kizhaev, and
  Chizhov]{Petrov1972}
M.~P. Petrov, G.~A. Smolenskii, A.~P. Paugurt, S.~A. Kizhaev, and M.~K.
  Chizhov.
\newblock Nuclear magnetic resonance and weak ferromagnetism in
  \mbox{FeBO$_3$}.
\newblock \emph{Solid State Physics}, 14:\penalty0 109--113, 2021.

\bibitem[Zvezdin et~al.(2021)Zvezdin, Kimel, Plokhov, and Zvezdin]{Zvezdin2020}
A.~K. Zvezdin, A.~V. Kimel, D.~I. Plokhov, and K.~A. Zvezdin.
\newblock Ultrafast spin dynamics in the iron borate easy-plane weak
  ferromagnet.
\newblock \emph{Journal of Experimental and Theoretical Physics}, 131:\penalty0
  130--138, 2021.

\bibitem[Kadomtseva et~al.(1972)Kadomtseva, Levitin, Popov, Seleznev, and
  Uskov]{Kadomtseva1972}
A.~M. Kadomtseva, R.~Z. Levitin, Yu.~F. Popov, V.~N. Seleznev, and V.~V. Uskov.
\newblock Magnetic and magnetoelastic properties of \mbox{FeBO$_3$}
  monocrystal.
\newblock \emph{Sov. Phys. Solid State}, 14:\penalty0 172, 1972.

\bibitem[Yagupov et~al.(2018)Yagupov, Strugatsky, Seleznyova, Mogilenec,
  Snegirev, Marchenkov, Kulikov, Eliovich, Frolov, Ogarkova, and
  Lyubutin]{Yagupov2018}
Sergey Yagupov, Mark Strugatsky, Kira Seleznyova, Yuliya Mogilenec, Nikita
  Snegirev, Nikita~V. Marchenkov, Anton~G. Kulikov, Yan~A. Eliovich, Kirill~V.
  Frolov, Yulia~L. Ogarkova, and Igor~S. Lyubutin.
\newblock Development of a synthesis technique and characterization of
  high-quality iron borate \mbox{FeBO$_3$} single crystals for applications in
  synchrotron technologies of a new generation.
\newblock \emph{Crystal Growth Design}, 18\penalty0 (12):\penalty0 7435--7440,
  October 2018.
\newblock URL \url{https://doi.org/10.1021/acs.cgd.8b01128}.

\bibitem[Mashkovich et~al.(2019)Mashkovich, Grishunin, Mikhaylovskiy, Zvezdin,
  Pisarev, Strugatsky, Christianen, Rasing, and Kimel]{Mashkovich2019}
E.~A. Mashkovich, K.~A. Grishunin, R.~V. Mikhaylovskiy, A.~K. Zvezdin, R.~V.
  Pisarev, M.~B. Strugatsky, P.~C.~M. Christianen, Th. Rasing, and A.~V. Kimel.
\newblock Terahertz optomagnetism: Nonlinear \mbox{THz} excitation of
  \mbox{GHz} spin waves in antiferromagnetic \mbox{FeBO$_3$}.
\newblock \emph{Physical Review Letters}, 123, 2019.

\bibitem[Ozhogin and Preobrazhenskii(1977)]{ozhogin1977effective}
VI~Ozhogin and VL~Preobrazhenskii.
\newblock Effective anharmonicity of elastic subsystem of antiferromagnets.
\newblock \emph{Sov.Phys. JETP}, 73:\penalty0 988--1000, 1977.

\bibitem[Zvezdin(1979)]{zvezdin1979dynamics}
AK~Zvezdin.
\newblock Dynamics of domain walls in weak ferromagnets.
\newblock \emph{ZhETF Pisma Redaktsiiu}, 29:\penalty0 605, 1979.

\bibitem[Bar'Yakhtar and Ivanov(1980)]{bar1980nonlinear}
IV~Bar'Yakhtar and BA~Ivanov.
\newblock Nonlinear waves in antiferromagnets.
\newblock \emph{Solid State Communications}, 34\penalty0 (7):\penalty0
  545--547, 1980.

\bibitem[Andreev and Marchenko(1980)]{Andreev1980}
Aleksandr~F Andreev and Vladimir~I Marchenko.
\newblock Symmetry and the macroscopic dynamics of magnetic materials.
\newblock \emph{Soviet Physics Uspekhi}, 23\penalty0 (1):\penalty0 21--34,
  January 1980.
\newblock \doi{10.1070/pu1980v023n01abeh004859}.
\newblock URL \url{https://doi.org/10.1070/pu1980v023n01abeh004859}.

\bibitem[Cheng et~al.(2014)Cheng, Xiao, Niu, and Brataas]{Cheng2014}
Ran Cheng, Jiang Xiao, Qian Niu, and Arne Brataas.
\newblock Spin pumping and spin-transfer torques in antiferromagnets.
\newblock \emph{Physical Review Letters}, 113\penalty0 (5), July 2014.
\newblock \doi{10.1103/physrevlett.113.057601}.
\newblock URL \url{https://doi.org/10.1103/physrevlett.113.057601}.

\bibitem[Ozhogin and Preobrazhensky(1988)]{Ozhogin1988}
V.~I. Ozhogin and V.~L. Preobrazhensky.
\newblock Anharmonicity of mixed modes and giant acoustic nonlinearity of
  antiferromagnetics.
\newblock \emph{Soviet Physics Uspekhi}, 31:\penalty0 713--729, 1988.

\bibitem[Morrish(1995)]{Morrish1995}
A~H Morrish.
\newblock \emph{Canted Antiferromagnetism: Hematite}.
\newblock {WORLD} {SCIENTIFIC}, January 1995.

\bibitem[Popov et~al.(2015)Popov, Zavislyak, Chumak, Strugatsky, Yagupov, and
  Srinivasan]{Popov2015}
M.~A. Popov, I.~V. Zavislyak, H.~L. Chumak, M.~B. Strugatsky, S.~V. Yagupov,
  and G.~Srinivasan.
\newblock Ferromagnetic resonance in a single crystal of iron borate and
  magnetic field tuning of hybrid oscillations in a composite structure with a
  dielectric: Experiment and theory.
\newblock \emph{Journal of Applied Physics}, 118\penalty0 (1):\penalty0 013903,
  July 2015.

\end{thebibliography}

\end{document}


\preprint{APS/123-QED}

\title{Supplementary Material to\\
Microwave Spin-Pumping from an Antiferromagnet FeBO$_3$}
\author{D.A.Gabrielyan}
\thanks{Authors to whom correspondence should be addressed:\\ \url{davidgabrielyan1997@gmail.com}}
\affiliation{
	Kotel'nikov Institute of Radioengineering and Electronics, Russian Academy of Sciences, Moscow 125009, Russia
}%
\affiliation{
	Moscow Power Engineering Institute, Moscow 111250, Russia
}%
\author{D.A. Volkov}
\affiliation{
	Kotel'nikov Institute of Radioengineering and Electronics, Russian Academy of Sciences, Moscow 125009, Russia
}%
\affiliation{
	Moscow Power Engineering Institute, Moscow 111250, Russia
}%
\author{E.E. Kozlova}
\thanks{\url{elizabethkozlova1@gmail.com}}
\affiliation{
	Kotel'nikov Institute of Radioengineering and Electronics, Russian Academy of Sciences, Moscow 125009, Russia
}%
\affiliation{
	Moscow Institute of Physics and Technology, Dolgoprudny 141701, Russia
}%
\author{A.R. Safin}
\affiliation{
	Kotel'nikov Institute of Radioengineering and Electronics, Russian Academy of Sciences, Moscow 125009, Russia
}%
\affiliation{
	Moscow Power Engineering Institute, Moscow 111250, Russia
}%
\author{D.V. Kalyabin}
\affiliation{
	Kotel'nikov Institute of Radioengineering and Electronics, Russian Academy of Sciences, Moscow 125009, Russia
}%
\affiliation{
	HSE University, Moscow, 101000, Russia
}%
\author{A.A. Klimov}
\affiliation{
	Kotel'nikov Institute of Radioengineering and Electronics, Russian Academy of Sciences, Moscow 125009, Russia
}%
\affiliation{
	MIREA - Russian Technological University, Moscow 119454, Russia
}%
\author{\\ V.L. Preobrazhensky}
\affiliation{
	Prokhorov General Physics Institute RAS, Moscow 119991, Russia
}%
\author{M.B. Strugatsky}
\affiliation{
	V.I. Vernadsky Crimean Federal University, Simferopol 295007, Russia
}%
\author{S.V. Yagupov}
\affiliation{
	V.I. Vernadsky Crimean Federal University, Simferopol 295007, Russia
}%
\author{I.E. Moskal}
\affiliation{
	Kotel'nikov Institute of Radioengineering and Electronics, Russian Academy of Sciences, Moscow 125009, Russia
}%
\author{G.A. Ovsyannikov}
\affiliation{
	Kotel'nikov Institute of Radioengineering and Electronics, Russian Academy of Sciences, Moscow 125009, Russia
}%
\author{S.A. Nikitov}
\affiliation{
	Kotel'nikov Institute of Radioengineering and Electronics, Russian Academy of Sciences, Moscow 125009, Russia
}%
\affiliation{
	Moscow Institute of Physics and Technology, Dolgoprudny 141701, Russia
}%
\affiliation{
	Saratov State University, Saratov 410012, Russia
}
\maketitle

\section{Static magnetic characteristics of the sample}
\begin{figure*}[!ht]  
\includegraphics[width=0.5\linewidth]{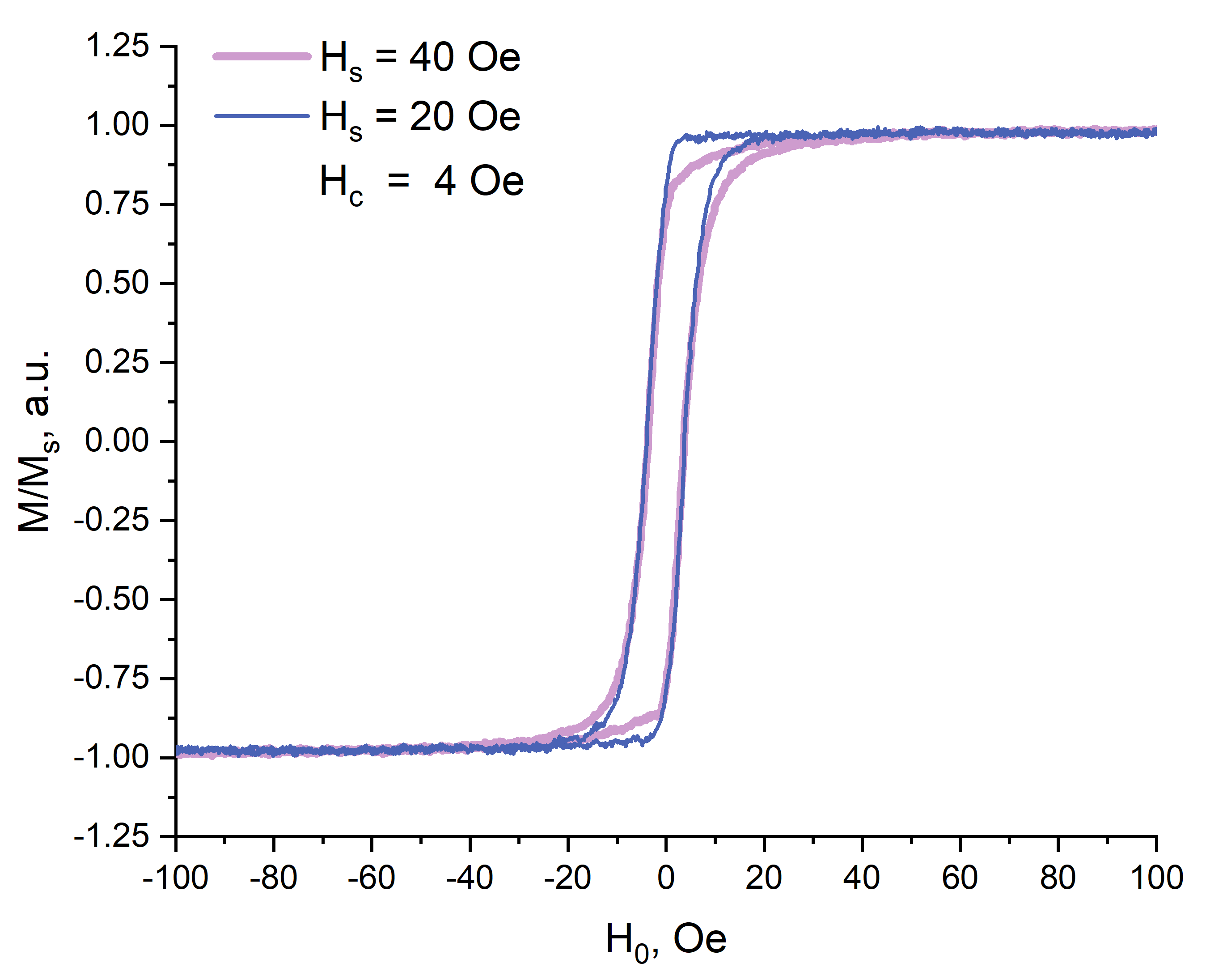}
\caption{
Hysteresis loops for two magnetization directions of FeBO$_3$ sample, curves are obtained for magnetization directions with minimum - 20 Oe and maximum - 40 Oe saturation fields H$_s$.
}
\label{fig:01}
\end{figure*}

The static magnetic characteristics of the FeBO$_3$ sample were obtained on a magneto-optical longitudinal Kerr effect (MOKE) setup. All measurements were performed in the magnetizing field H$_0$ lying in the sample plane. The type of hysteresis loops and the magnitude of the MOKE response depend on the direction of the applied field. When the sample was rotated around the normal to its surface, hysteresis loops were recorded (Fig.~\ref{fig:01}). The magnitude of the coercive force H$_c$ in all measurements did not exceed 4 Oe. The value of the saturation field H$_s$ varied from 20 Oe to 40 Oe, which confirms the presence of sixth-order anisotropy in the FeBO$_3$ sample.

\section{Theoretical model}
To describe the dynamics of the antiferromagnet considered in the Letter, we use equations considered in  \cite{Ozhogin1988} due to their simplicity in the analysis of both linear and nonlinear dynamic effects.

We examine the output expression for ISHE voltage in more depth. For this purpose, let us represent the N\'eel vector as the sum of the static component $\mathbf{l}_0$, describing the AFM ground state, which is oriented along the $\mathbf{e}_2$ axis, and the small dynamic vector $\mathbf{s}(t)$:
\begin{equation}\label{eq:II.1}
    \mathbf{l}(t)=\mathbf{l}_0+\mathbf{s}(t).
    \tag{II.1}
\end{equation}

The vectors $\mathbf{l}_0$ and $\mathbf{s}(t)$ satisfy the orthogonality condition, that is, $(\mathbf{l}_0\cdot \mathbf{s})=0$.  

The equation for the small-amplitudes dynamics of the N\'eel vector is obtained by substitution Eq.(1) in Eq.(2) of the main text:
\begin{equation}\label{eq:02}
\begin{array}{c}
    \frac{d^2\mathbf{s}}{dt^2}+\gamma_{\tiny{\textrm{eff}}}\frac{d\mathbf{s}}{dt}+[\widehat{\Omega}-(\mathbf{l}_0\cdot\widehat{\Omega}\mathbf{l}_0)\widehat{I}]\mathbf{s}=-\widehat{\Theta}\cdot\mathbf{x}\cdot e^{i \omega t},
    \end{array}
    \tag{II.2}
\end{equation}
where matrices $\widehat{\Omega}, \widehat{I}, \widehat{\Theta}$ can be represented as 
\begin{equation}\label{eq:03}
\begin{array}{c}
    \widehat{\Omega}=\left(\omega_{\tiny{\textrm{H}}}^2+\omega_{\tiny{\textrm{H}}}\omega_{\tiny{\textrm{DMI}}}+\omega_{\tiny{\textrm{ex}}}\omega_{\tiny{\textrm{me}}}\right)\mathbf{x}\otimes\mathbf{x}+\left(\omega_{\tiny{\textrm{DMI}}}^2+\omega_{\tiny{\textrm{H}}}\omega_{\tiny{\textrm{DMI}}}+\omega_{\tiny{\textrm{ex}}}\omega_{\tiny{\textrm{a}}}\right)\mathbf{z}\otimes\mathbf{z},\\
    \\
    \widehat{I}=\textrm{diag}\left(1,1,1\right), \widehat{\Theta}=\omega_{\tiny{\textrm{H}}}\gamma h_{\tiny{\textrm{RF}}}\mathbf{x}\otimes\mathbf{x}.
    \end{array}\tag{II.3}
    \end{equation}

To obtain an expression for ISHE voltage of the low-frequency mode, let us present a dynamic vector $\mathbf{s}(t)$ in the form of:
\begin{equation}\label{eq:04}
\begin{array}{c}
    \mathbf{s}\left(t\right) = A\left(\omega\right)\sin\left(\omega t + \phi \left(\omega\right)\right)\mathbf{x},
    \end{array}\tag{II.4}
    \end{equation}
where A is an amplitude, $\phi$ is a phase depending on the frequency of the input signal.

By combining equations (II.2) and (II.4), we get expressions for the amplitude and the phase:
\begin{equation}\label{eq:05}
\begin{array}{c}
    A\left(\omega\right) = \frac{\omega_{\tiny{\textrm{DMI}}}\omega_{\tiny{\textrm{RF}}}}{\sqrt{\left(\omega^2-\omega_0^2\right)^2+\left(\gamma_{\tiny{\textrm{eff}}}\omega\right)^2}},
    \end{array}\tag{II.5}
    \end{equation}
\begin{equation}\label{eq:06}
\begin{array}{c}
    \tan \phi\left(\omega\right) = \frac{\gamma_{\tiny{\textrm{eff}}}\omega}{\omega^2 - \omega_0^2}.
    \end{array}\tag{II.6}
    \end{equation}

Substitution of the solution Eq.(II.4,5)
in Eq. (1) and Eq.(6) of the main text gives in an explicit way an expression for ISHE
voltage of the low-frequency mode:
\begin{equation}\label{eq:07}
\begin{array}{c}
    V_{\tiny{\textrm{out}}}\left(\omega\right) = \kappa\cdot\frac{\left(\omega_{\tiny{\textrm{H}}}+\omega_{\tiny{\textrm{DMI}}}\right)\left(\omega_{\tiny{\textrm{DMI}}}\omega_{\tiny{\textrm{RF}}}\right)^2}{\left(\omega^2-\omega_{\tiny{\textrm{QFMR}}}^2\right)^2+\left(\gamma_{\tiny{\textrm{eff}}}\omega\right)^2}\cdot\omega^2,
    \end{array}\tag{II.7}
    \end{equation}
where $\omega_{\tiny{\textrm{RF}}}= \gamma\cdot h_{\tiny{\textrm{RF}}}$. We note that for the quasi-ferromagnetic mode only the second term in the Eq.(6) contributes to the ISHE voltage (II.7).

\section{VNA-FMR experimental setup}

\begin{figure*}[!ht]  
\includegraphics[width=0.5\linewidth]{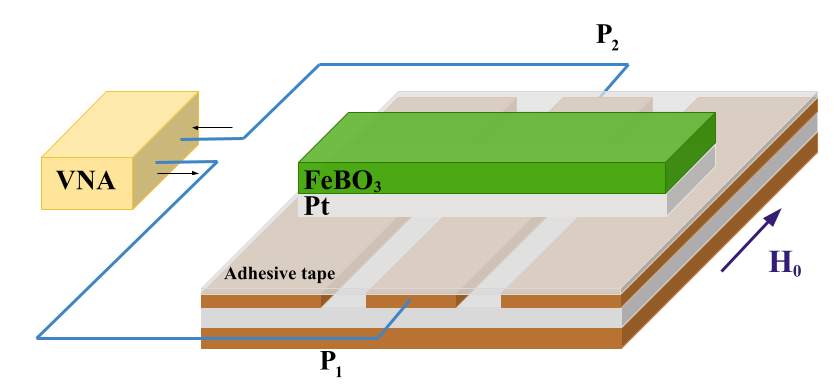}
\caption{
The \textit{VNA-FMR} setup.
}
\label{fig:02}
\end{figure*}

The $\textrm{S}_{21}$-parameter of the coplanar waveguide on which the sample is placed in the external magnetic field and measured using the \textit{VNA-FMR} method (Fig.~\ref{fig:02}). $\textrm{S}_{21}$-parameter shows the ratio of the output power (P$_2$) of the coplanar waveguide to the input power (P$_1$). 

\begin{figure*}[!ht]  
\includegraphics[width=1\linewidth]{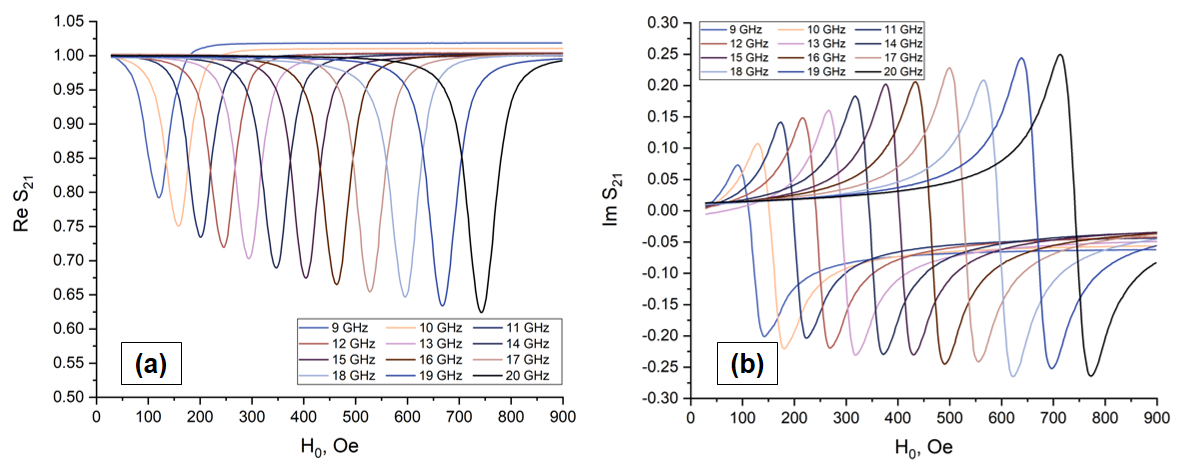}
\caption{
FMR spectra of FeBO$_3$/Pt heterostructure measured by the \textit{VNA-FMR} method. (a) Real part of FMR spectra. (b) Imaginary part of FMR spectra.
}
\label{fig:03}
\end{figure*}

The Fig.~\ref{fig:03} shows the real and imaginary part of the ferromagnetic resonance spectrum, which gives information about the absorption intensity of microwave power and the signal phase changing. 
The following expressions are used to determine the magnitude and phase of the spectrum of the real and imaginary parts of $\textrm{S}_{21}$-parameter:

\begin{equation}\label{eq:08}
\begin{array}{c}
    \textrm{mag}\left(\textrm{S}_{21}\right)=\sqrt{\left(\textrm{Re}\left(\textrm{S}_{21}\right)\right)^2+\left(\textrm{Im}\left(\textrm{S}_{21}\right)\right)^2},
    \end{array}\tag{III.1}
    \end{equation}
\begin{equation}\label{eq:09}
\begin{array}{c}
    \textrm{phase}\left(\textrm{S}_{21}\right)= 
    \begin{cases}
        \textrm{arctg}\left(\frac{\textrm{Im}\left(\textrm{S}_{21}\right)}{\textrm{Re}\left(\textrm{S}_{21}\right)}\right) &\text{if $\textrm{Re}\left(\textrm{S}_{21}\right) \geq 0$}\\
        \pi + \textrm{arctg}\left(\frac{\textrm{Im}\left(\textrm{S}_{21}\right)}{\textrm{Re}\left(\textrm{S}_{21}\right)}\right) &\text{if $\textrm{Re}\left(\textrm{S}_{21}\right) < 0$}
    \end{cases}.
    \end{array}\tag{III.2}
    \end{equation}

\bibliography{main}